# WHY THE STANDARD DATA PROCESSING SHOULD BE CHANGED

Yefim Bakman

Tel-Aviv University, *E-mail:* bakman@post.tau.ac.il

**Abstract**. The basic statistical methods of data representation did not change since their emergence. Their simplicity was dictated by the intricacies of computations in the before computers epoch. It turns out that such approach is not uniquely possible in the presence of quick computers. The suggested here method improves significantly the reliability of data processing and their graphical representation.

In this paper we show problems of the standard data processing which can bring to incorrect results. A method solving these problems is proposed. It is based on modification of data representation. The method was implemented in a computer program Consensus5. The program performances are illustrated through varied examples.

Key words: data processing, fuzzy statistics, graphical methods, erroneous data detecting, faulty measurers, robust estimators, students' estimations, unskilled experts.

## 1. Data representation by means of fuzzy numbers

It is frequently said that statistics is an accurate science about non-accurate things. Indeed there exists incompatibility between measurements and the statistical method of their description. If measurements are inaccurate, the method for their representation must include the measurement errors as their integral part. It is this problem that leads to distorted graphical data representation by histogram.

In order to correct the situation, we will refer data as fuzzy numbers (see, for example, Klir [1], Buckley [2]). According to this method a measured value is represented by means of a membership function which most often has the trapezoidal shape (see Fig.1).

If $A(x)$ is a membership function of some measured value V, then $A(x)$ is the confidence that V=x. We can see from Fig.1 that $A(7.4) = 1$ just like $A(7.3) = 1$, that is deviation from measured value 7.3 within the limits of the measurement error does not alter the confidence. However $A(7.45) = 0.5$ and then the confidence gradually falls till zero at x=7.5 remaining continuous all the time.

When every value of the sample is presented in the form of a fuzzy number, one can built the common membership function. Its graph will be the new graphic representation of the data. This method of graph plotting was implemented in the computer program Consensus5.

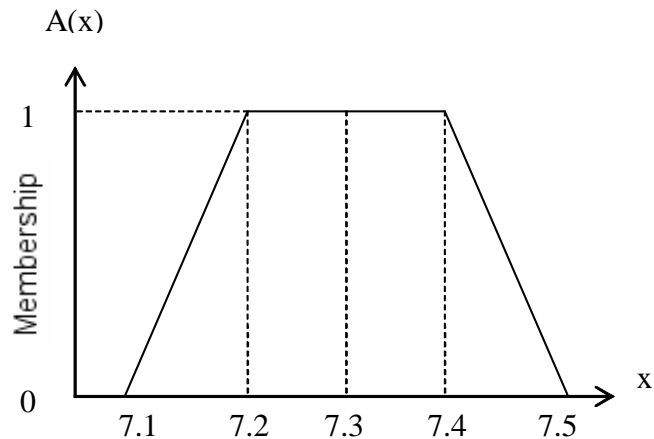

Fig.1. A membership function example.

As an example I took 250 random numbers generated according to the normal distribution N(0,3). The program Consensus5 gives ideal approximation to the normal distribution for these data (see Fig.2a), whereas the histogram is far from the normal curve (Fig.2b).

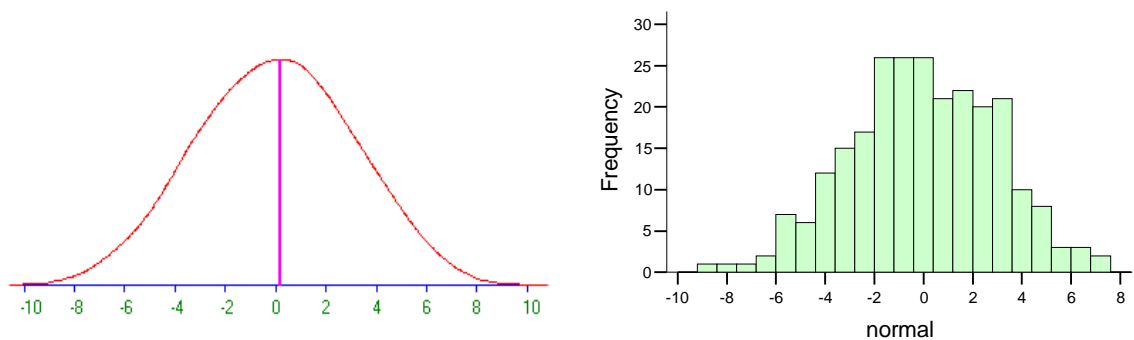

Fig.2. The graphic representation of 250 normally distributed random numbers built by the computer program Consensus5 (left), and the conventional histogram (right).

As another example we will consider a plot chart in which one variable depends on the other. The dependent variable is the number of building permissions given to applicants during a certain month, that is to say the independent variable is time. The plot chart is shown in Fig.3. The possibility of mistaken dating is not taken into account in this chart, for example, the applicants postponed their requests in vacation periods or because of bad weather which could lead to the graph fall in one month and the following leap in the next month. Such jumps of the graph curve hinder the tendency.

The same data were fed to the program Consensus5, and the resulting chart is shown in Fig.4. Because of the fuzzy numbers representation, transfer of permissions to the neighboring month does not change the graph. Instead of abrupt changes one may concentrate on the tendency.

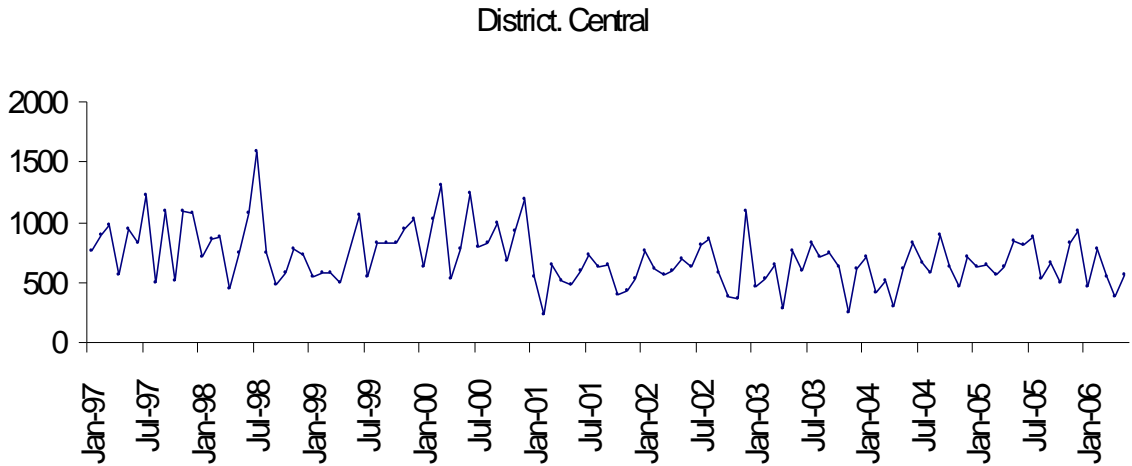

Fig.3. Each point of the plot chart shows the number of building permissions given to applicants in a certain month (total 113 points).

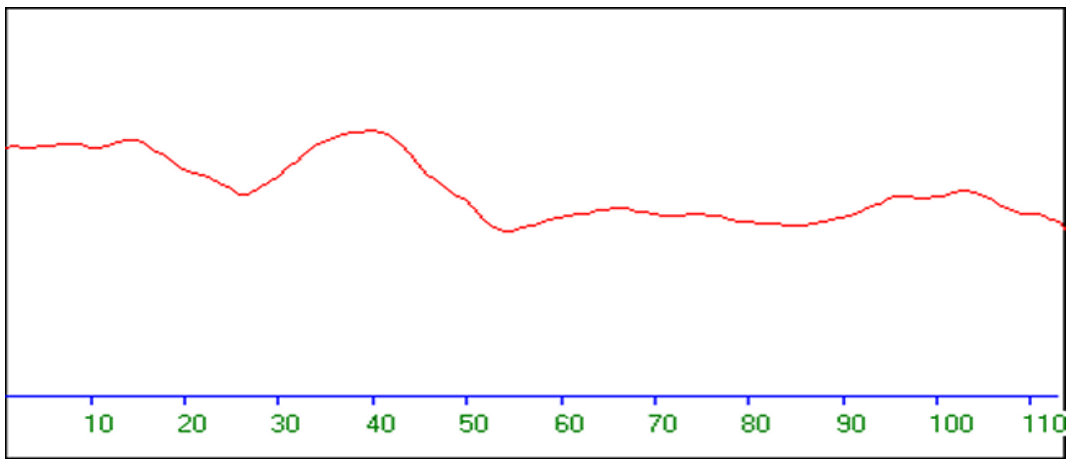

Fig.4. The same data as in Fig.3 represented by means of fuzzy numbers show the general tendency instead of local fluctuations.

According to Fig.3 the maximal number of permissions was in July 1998, but this peak turned out to be false, in fact this period did not differ from the preceding 1997 year.

## 2. Consensus and erroneous data detection

I was a witness of a real event, occurred at statistics lesson. The teacher handed out survey forms to his students. The students were to estimate the level of his teaching, the degree of preparation to lessons, the availability to answer questions, and so on, totally 12 estimates.

One of the students was new. He, that was not present even at one lesson, with serious expression on his face, began to mark grades in the form. The teacher asked him, how he can estimate his teaching, if he had not attended even one of the lessons. The student thought for a moment and then said that the teacher was right, and put the form aside. This event caused me to start development of the new approach.

That was an abnormal case, of course, but how many students unable to estimate their teachers and nevertheless estimate them. A method for such "experts" detection is needed. Nowadays for detection of erroneous values a standard procedure is used, in which the extreme values are considered erroneous. In effect not always extreme values are erroneous and not always erroneous values are extreme ones. We will show this with the help of the following example.

Let two magnitudes X and Y were measured by means of three properly working measurers S1-S3 with measuring error being 0.2. The obtained values are presented in Table 1.

|   | S1  | S2 | S3  |
|---|-----|----|-----|
| X | 1.9 | 2  | 2.1 |
| Y | 0.9 | 1  | 1.1 |

Table 1. Three properly measured values of the magnitudes X and Y.

For these data we shall calculate main statistical indicators, and then we shall verify their robustness by means of adding to the table supplementary data received from three defected measurers S4-S6.

|   | S4 | S5 | S6 |
|---|----|----|----|
| X | 4  | 6  | 7  |
| Y | 3  | 5  | 4  |

Table.2. Three erroneous values of the magnitudes X and Y.

As an alternative indicator we shall consider consensus, whose essence can be better expressed with the help of graphic representation. In Fig.5 the six measured values S1-S6 are shown in the form of squares denoting the borders of permitted values taking into account the measurement error. As it can be seen from Fig.5, squares S1-S3 possess common zone in the vicinity of a point (2; 1) - at this point the indicated squares overlap one another.

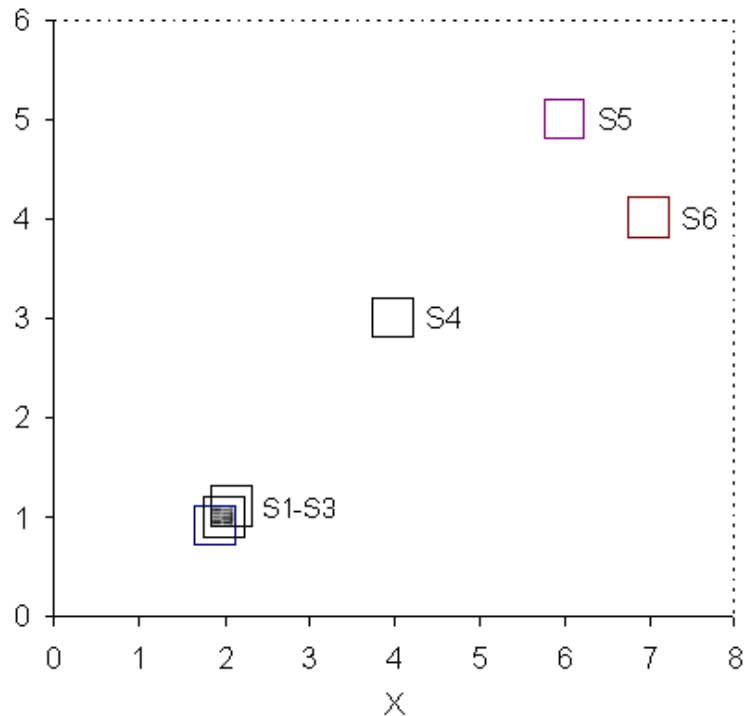

Fig.5. Values S1-S6 are presented in the form of squares marking the borders of values within the measurement error. The zone of consensus (where the squares S1-S3 overlap) is painted.

**Definition**: *Consensus* is a point or zone, at which the greatest number of measured values overlap (taking into account their measurement errors).

Mode presents a particular case of consensus, when measurements are absolutely accurate (null error). If all the values of our example were absolutely accurate, then instead of the squares there would be points in Fig.5 and no points would overlap.

The idea of consensus is as follows: correct values are close to the true value, consequently, they are close to each other and their squares overlap forming consensus. On the contrary, erroneous values are scattered around randomly. They are far from each other having low chances to overlap; therefore they cannot change the consensus. This means that consensus is robust against adding erroneous data; it also helps to reveal erroneous values. Fig.5 shows that the measurements S4-S6 are out of consensus; consequently, they are erroneous and should be expelled from the consideration.

Now we will verify the robustness of standard statistical estimators against adding erroneous data.

| Estimator | X value before and after adding the erroneous data | | | Y value before and after adding the erroneous data | | |
|---|---|---|---|---|---|---|
| | before | after | deviation | before | after | deviation |
| consensus | 2 | 2 | **0** | 1 | 1 | **0** |
| mean | 2 | 3.83 | 1.83 | 1 | 2.50 | 1.50 |
| median | 2 | 3.05 | 1.05 | 1 | 2.05 | 1.05 |

Table 3. Comparison of the changes of basic statistical estimators for X and Y as a result of entering the erroneous measurements S4-S6.

As it is seen in Table 3, after entering the erroneous measurements the consensus did not change, whereas the sample mean and median changed drastically that evidences their instability. The changes of M-estimators are discussed in Appendix.

## 3. Computer program Consensus5 testing

In addition the program was tested on two surveys. The respondents answered 9 questions (estimate on the scale 1 to 7 the climate of Israel, the public health system, the economic status and so on).

After the consensus was calculated for every question, four virtual respondents with random estimates were added to the survey. The program evaluated them as incompetent, and the calculated consensus coincided with the first run value (see Fig.6). Thus the consensus proved its robustness.

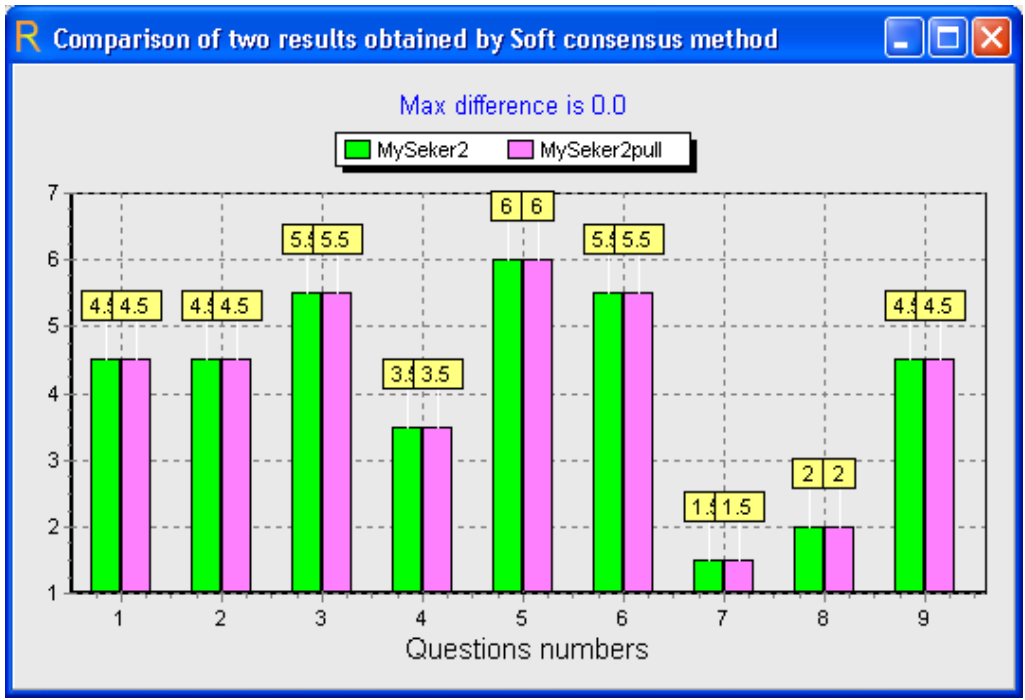

Fig.6. After addition of four artificial respondents with random estimates to the survey the program Consensus5 labeled them as erroneous and deleted them from the data. It is natural that the consensus remained unchanged.

The sample means and medians were also calculated for the survey (see Fig.7). The greatest difference of the sample means between the two runs was 1.2 points, whereas for the medians the maximum difference was 3 points which evidences instability of these two parameters against adding erroneous data to the survey.

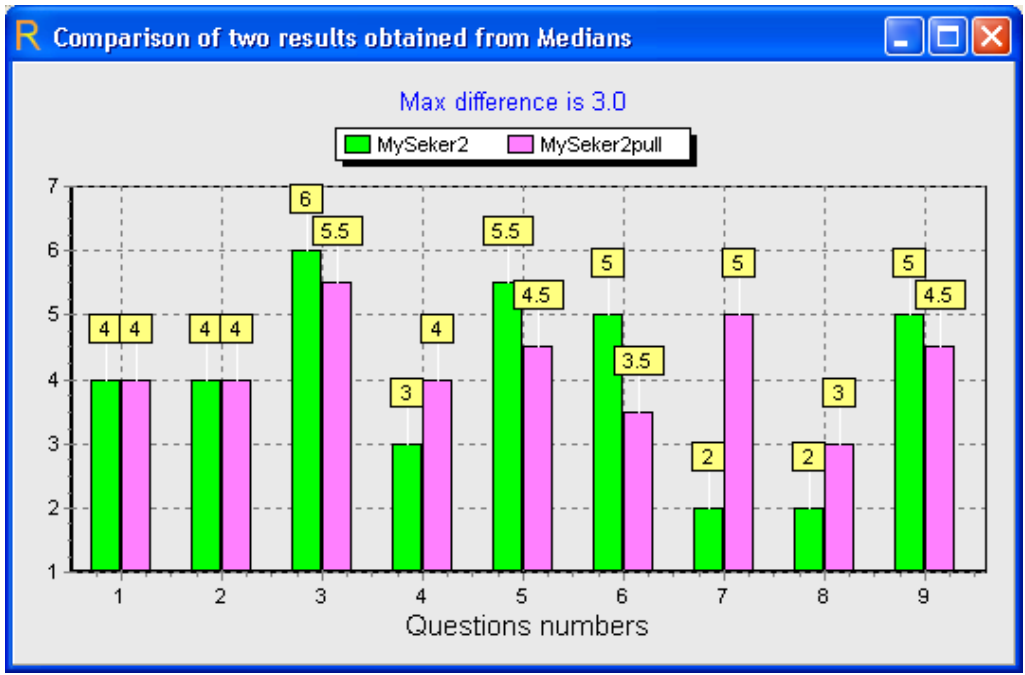

Fig.7. After addition of four artificial respondents with random estimates to the survey the greatest difference of the sample medians composed 3 points. This evidences that the median is not robust against adding erroneous data to the survey.

The testing of the computer program Consensus5 demonstrated following advantages of the suggested method for data processing:
 a) Felicitous graphical representation even for small samples.
 b) Consensus is robust against erroneous measurements.
 c) Consensus computation helps to detect faulty measuring devices and unskilled experts.

Conclusion

All the bundle of problems associated with the standard statistical approach was solved with the help of one unique correction of the method for data representation. In the light of the suggested correction the old method appears to be a particular case, in which the interval of possible values degenerates into a point, herewith the consensus turns into the mode. Such representation would be true if all measurements were absolutely accurate which is unreal.

It is most probable that the choice of the oversimplified method for data representation was dictated by the computational difficulties in the before computers epoch. Nowadays we can afford the luxury of the choice of the pertinent algorithm for data processing not being restricted by number of computations.

I invite everyone to send me his/her data to test the program Consensus5. Please write the word "Consensus" as the subject of the electronic message.

*References*

[1] G. Klir and B. Yuan. *Fuzzy Sets and Fuzzy Logic, Theory and Applications*, Prentice Hall, Englewood Cliffs, NJ, 1995.
[2] J. Buckley. *Fuzzy Statistics*, NY, 2004.

## Appendix

The definition of M-estimators says that they are robust alternatives to the sample mean and median. Contrary to this definition their deviations were grater than those of the median (see Table 4).

| M-estimators | X value before and after adding the erroneous data | | | Y value before and after adding the erroneous data | | |
|---|---|---|---|---|---|---|
| | before | after | deviation | before | after | deviation |
| Huber's | 2 | 3.23 | 1.23 | 1 | 2.23 | 1.23 |
| Tukey's | 2 | 3.05 | 1.05 | 1 | 2.30 | 1.30 |
| Hampel's | 2 | 3.43 | 1.43 | 1 | 2.37 | 1.37 |
| Andrews' | 2 | 3.05 | 1.05 | 1 | 2.31 | 1.31 |

Table 4. Comparison of M-estimators changes for X and Y as a result of adding the erroneous measurements S4-S6 to the sample.